\documentclass[conference]{IEEEtran}
\IEEEoverridecommandlockouts

\usepackage{cite}
\usepackage{amsmath,amssymb,amsfonts}
\usepackage{algorithmic}
\usepackage{graphicx}
\usepackage{textcomp}
\usepackage{xcolor}
\usepackage{amsthm}
\newtheorem{definition}{Definition}

\def\BibTeX{{\rm B\kern-.05em{\sc i\kern-.025em b}\kern-.08em
    T\kern-.1667em\lower.7ex\hbox{E}\kern-.125emX}}
\begin{document}

\title{DyLoC: A Dual-Layer Architecture for Secure and Trainable Quantum Machine Learning Under Polynomial-DLA constraint\\
\thanks{This work was supported in part by the National Natural Science Foundation of China under Grant  62471020 and in part by the Beijing Natural Science Foundation under Grant L251066}
}

\author{\IEEEauthorblockN{ Chenyi Zhang}
\IEEEauthorblockA{\textit{School of Cyber Science and Technology} \\
\textit{Beihang University}\\
100083, Beijing, China \\
zhangchenyi@buaa.edu.cn}
\and
\IEEEauthorblockN{Tao Shang \textsuperscript{*}  \thanks{\textsuperscript{*} Corresponding author: Tao Shang, email: shangtao@buaa.edu.cn}}
\IEEEauthorblockA{\textit{School of Cyber Science and Technology} \\
\textit{Beihang University}\\
100083, Beijing, China \\
shangtao@buaa.edu.cn}
\and
\IEEEauthorblockN{Chao Guo}
\IEEEauthorblockA{\textit{School of Cyber Science and Technology} \\
\textit{Beihang University}\\
100083, Beijing, China \\
guochao2539@buaa.edu.cn}
\and
\IEEEauthorblockN{Ruohan He}
\IEEEauthorblockA{\textit{School of Cyber Science and Technology} \\
\textit{Beihang University}\\
100083, Beijing, China \\
SY2539116@buaa.edu.cn}
}

\maketitle

\begin{abstract}
Variational quantum circuits face a critical trade-off between privacy and trainability. High expressivity required for robust privacy induces exponentially large dynamical Lie algebras. This structure inevitably leads to barren plateaus. Conversely, trainable models restricted to polynomial-sized algebras remain transparent to algebraic attacks. To resolve this impasse, DyLoC is proposed. This dual-layer architecture employs an orthogonal decoupling strategy. Trainability is anchored to a polynomial-DLA ansatz while privacy is externalized to the input and output interfaces. Specifically, Truncated Chebyshev Graph Encoding (TCGE) is employed to thwart snapshot inversion. Dynamic Local Scrambling (DLS) is utilized to obfuscate gradients. Experiments demonstrate that DyLoC maintains baseline-level convergence with a final loss of 0.186. It outperforms the baseline by increasing the gradient reconstruction error by 13 orders of magnitude. Furthermore, snapshot inversion attacks are blocked when the reconstruction mean squared error exceeds 2.0. These results confirm that DyLoC effectively establishes a verifiable pathway for secure and trainable quantum machine learning.

\end{abstract}


\section{Introduction}
Quantum machine learning (QML) \cite{biamonte2017quantum} has emerged as a key application for noisy intermediate-scale quantum (NISQ) devices \cite{preskill2018quantum}. Variational Quantum Circuits (VQC) \cite{cerezo2021variational}, serving as the backbone of NISQ-era QML, are increasingly deployed in sensitive domains such as finance and healthcare. In these architectures, classical data is encoded into quantum states and processed via parameterized ansatzes. The optimization relies on classical feedback loops based on measurement gradients. Consequently, data privacy has become a key concern.

Unique privacy vulnerabilities exist in quantum architectures. Recent theoretical advancements have characterized these risks within the Lie algebra supported ansatz (LASA) framework \cite{heredge2025characterizing}. Research indicates a linear dependency between training gradients and the "snapshots" of encoded quantum states. Such algebraic dependency enables adversaries to reconstruct intermediate quantum states from public gradients via algebraic means, which constitutes a Weak Privacy Breach. Subsequently, original input data can be mathematically inverted from these snapshots, constituting a Strong Privacy Breach.

Addressing these vulnerabilities presents a fundamental challenge known as the "privacy-trainability trade-off." According to the dynamical Lie algebra (DLA) theory \cite{qvarfort2025solving}, robust privacy typically necessitates circuits with high expressivity, which corresponds to an exponentially large DLA dimension. Such immense algebraic dimensions inevitably lead to the barren plateau (BP) phenomenon, rendering the model untrainable \cite{mcclean2018barren,cerezo2021cost,holmes2022connecting}. Conversely, VQC architectures designed for trainability are restricted to polynomial-sized DLAs. Heredge et al. \cite{heredge2025characterizing} prove that these trainable models are inherently transparent to algebraic attacks due to their low complexity.

Current defenses encounter challenges in resolving this contradiction effectively. Differential privacy limits utility, while cryptographic methods impose prohibitive resource overheads. To resolve this impasse, this paper proposes a dual-layer architecture based on the \textbf{Dy}namic \textbf{Lo}cal Scrambling and the truncated \textbf{C}hebyshev graph encoding (DyLoC). The core innovation of DyLoC involves an orthogonal decoupling strategy, which separates the source of privacy from the source of trainability. Unlike prior works relying on deepened circuits to hide information, trainability is anchored to a polynomial-DLA ansatz while privacy mechanisms are externalized to the input and output interfaces. The input interface employs the Truncated Chebyshev Graph Encoding (TCGE). TCGE utilizes a Chebyshev Tower strategy combined with Graph State initialization. The separability assumption required by known inversion algorithms is explicitly violated while a constant circuit depth is maintained to preserve signal variance. The output interface employs the dynamic local scrambling (DLS). DLS utilizes time-varying local random unitary transformations. The linear relationship between gradients and snapshots is obfuscated to prevent state recovery. Theoretical and experimental analyses confirm that the locality and shallow depth of DyLoC preserve the variance of the gradient signal.

The main contributions of our work are:
\begin{enumerate}
    \item \textbf{Proposal of an orthogonal decoupling strategy}: A theoretical framework is established to separate privacy protection from ansatz expressibility. By utilizing high-complexity input/output mappings, the architecture breaks the privacy-trainability trade-off inherent in traditional designs under polynomial DLA constraint.
    \item \textbf{Design of the DyLoC architecture}: A dual-layer defense system comprising TCGE and DLS is constructed. TCGE utilizes a tower-structure mapping and graph-state entanglement to thwart snapshot inversion (Strong Privacy), while DLS employs perturbative local unitaries to obfuscate gradient-snapshot linearity(Weak Privacy).
    \item \textbf{Demonstration of superior privacy-utility balance}: Verification through experiments shows that DyLoC maintains convergence rates comparable to unprotected baselines. The proposed scheme outperforms quantum differential privacy in both gradient reconstruction error and landscape ruggedness without incurring the utility loss associated with noise injection.
\end{enumerate}

\section{Related works}

\subsection{Trainability and barren plateaus} 
The scalability of variational algorithms is constrained by the barren plateau (BP) phenomenon \cite{mcclean2018barren,cerezo2021cost,holmes2022connecting}. Ragone et al. \cite{ragone2024lie} unified the origin of this phenomenon under the dynamical Lie algebra (DLA) framework. Theoretical proofs indicate that gradient variance is inversely proportional to the DLA dimension. Consequently, high-expressivity ansatzes that generate exponential DLAs are rendered untrainable. This theoretical bound forces a restriction to structured ansatzes with polynomial-level DLAs, such as the Hamiltonian variational ansatz \cite{wiersema2020exploring}.

\subsection{Limitations of existing defenses}

Current defenses face key trade-offs between security and utility. Within the field of quantum differential privacy \cite{watkins2023quantum,ju2024harnessing}, Du et al. \cite{du2021quantum} proposed the injection of noise into gradients to satisfy privacy bounds. Gong et al. [14] protect quantum learning systems from adversarial attacks by randomly encoding legitimate data samples. While theoretically sound, stochastic disruption inevitably degrades model utility and convergence stability.

Beyond noise injection, alternative strategies face implementation hurdles. Cryptographic protocols such as blind quantum computing \cite{broadbent2009universal,li2021quantum} offer secure delegation but impose communication overheads that render them impractical for iterative variational training tasks requiring frequent updates. Furthermore, although high-frequency encodings based on data re-uploading introduce nonlinearity \cite{kumar2023expressive}, standard implementations often retain separable structures. This characteristic leaves them vulnerable to subsystem-based inversion attacks identified in prior research \cite{heredge2025characterizing}.

In contrast, the DyLoC architecture proposed in this paper employs a shallow and entangled graph-state structure. This approach breaks the separability assumption without inducing volume-law entanglement. Consequently, the model is secured against algebraic attacks while trainability is preserved.

\section{DyLoC: A dual-layer defense architecture}
\subsection{Overview of DyLoC}
\subsubsection{Preliminaries}. Consider an $n$-qubit system with the Hilbert space $\mathcal{H} = (\mathbb{C}^2)^{\otimes n}$. A standard VQC model comprises a data encoding map $V(x)$ and a parameterized variational ansatz $U(\theta)$. Let $x \in \mathcal{X} \subset \mathbb{R}^d$ denote the input data. The encoding map prepares an input-dependent density state:
\begin{align}
    \rho(x) = V(x)|0\rangle\langle0|^{\otimes n}V(x)^\dagger
\end{align}

Subsequently, the state evolves under the variational circuit $U(\theta)$, typically defined as a sequence of Pauli rotations parameterized by $\theta \in \mathbb{R}^D$:
\begin{align}
    U(\theta) = \prod_{k=1}^D e^{-i\theta_k H_{\nu(k)}}
\end{align}

where $\theta = [\theta_1, \dots, \theta_D]$ are trainable parameters and $\{H_1, \dots, H_N\}$ constitutes a set of Hermitian generators. 

\begin{definition}[The Dynamical Lie Algebra (DLA)]
denoted as $\mathfrak{g}$, is defined as the real linear span of the nested commutators generated by $\{iH_1, \dots, iH_N\}$. The output of the model is the expectation value of an observable $O$:
\begin{align}
    y_\theta(x) = \text{Tr}(O U(\theta) \rho(x) U(\theta)^\dagger)
\end{align}
\end{definition}

To ensure trainability and avoid barren plateaus, the model is assumed to satisfy the Lie algebra supported ansatz (LASA) condition, where the measurement operator satisfies $iO\in\mathfrak{g}$. Furthermore, the dimension of the DLA is assumed to scale polynomially with the number of qubits, i.e., $dim(\mathfrak{g})=poly(n)$. Under these conditions, the output can be expressed as a linear contraction of a snapshot vector $e_{snap}(x)\in\mathbb{R}^{dim(\mathfrak{g})}$:

\begin{align}
   y_\theta(x) = \mu^T \text{Ad}_{U(\theta)} e_{snap}(x)
\end{align}

where $[e_{snap}]_\alpha = \text{Tr}(B_\alpha \rho(x))$ represents the projection of the input state onto the orthonormal basis $\{B_\alpha\}$ of $\mathfrak{g}$

\subsubsection{Thread model}. A white-box adversary aiming to reconstruct the private input training data $x$ from the shared gradients is considered. The attack consists of two sequential phases.

\begin{enumerate}
    \item \textbf{Weak Privacy Breach (Snapshot Recovery)}. The adversary observes the gradients $C_j = \partial y_\theta / \partial \theta_j$. Under the LASA condition, the gradient is linearly related to the snapshots:
    \begin{align}
        C_j = \sum_{\alpha=1}^{\dim(\mathfrak{g})} \Omega_{j\alpha}(\theta) [e_{snap}]_\alpha
    \end{align}
    
    Since $\theta$ and the ansatz structure are known, the matrix $\Omega$ is deterministic. If $\dim(\mathfrak{g})$ is polynomial, the adversary can efficiently solve this system to recover $e_{snap}$ using algorithms such as the Snapshot recovery algorithm described in \cite{heredge2025characterizing}.
    \item \textbf{Strong Privacy Breach (Snapshot Inversion)}. Upon recovering $e_{snap}$, the adversary attempts to invert the encoding map $\mathcal{M}: x \to e_{snap}(x)$ to obtain $x$. Prior work has demonstrated that, for standard encodings (e.g., Pauli product maps), snapshot inversion is feasible in polynomial time. This can be achieved using the Algorithm of Snapshot Inversion for General Pauli Encodings\cite{heredge2025characterizing}, provided the encoding state allows for separable subsystem analysis.
\end{enumerate}

\begin{figure}
    \centering
    \includegraphics[width=1\linewidth]{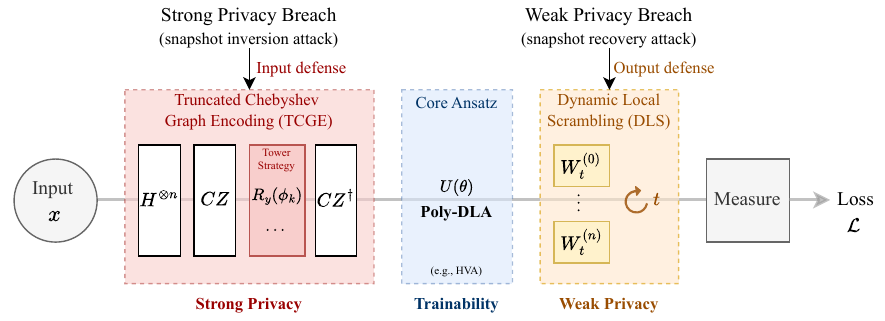}
    \caption{\textbf{DyLoC architecture.} The DyLoC architecture secures the trainable polynomial-DLA core by deploying TCGE at the input to create a rugged landscape against inversion attacks and DLS at the output to dynamically obfuscate gradients. This orthogonal design achieves dual-layer privacy protection while strictly preserving the gradient signal essential for model convergence.}
    \label{structure}
\end{figure}
As illustrated in Fig. \ref{structure}, the DyLoC architecture constructs a secure data flow by sandwiching a trainable core between two specialized defense layers. The processing pipeline begins with TCGE at the input stage. Here, a layer of Hadamard gates ($H^{\otimes n}$) and a constant-depth CZ ladder prepare a multipartite graph state, onto which data is encoded via a Chebyshev Tower strategy ($R_y(\phi_k)$). This structure enforces global entanglement and high nonlinearity to block strong privacy breaches. The quantum state then evolves through the Core Ansatz, which is structurally constrained to generate a polynomial dynamical Lie algebra (Poly-DLA), thereby guaranteeing trainability. Finally, at the output stage, DLS mechanism applies time-varying local random unitaries ($w_t$) to the state before measurement. This introduces stochasticity into the gradient observation channel, effectively thwarting weak privacy breaches based on snapshot recovery, without inducing measurement-dependent barren plateaus.

\subsection{Dynamic local scrambling (DLS)}

First, we identify the inherent vulnerability of static LASA systems. If the measurement operator $O$ is static, public, and satisfies the condition $iO \in \mathfrak{g}$, the gradient vector $\mathbf{C}$ satisfies a deterministic linear relationship $\mathbf{C} = \mathbf{A}_{static} \cdot e_{snap}$. Since the adversary possesses knowledge of the ansatz structure, the matrix $\mathbf{A}_{static}$ is fully computable. This reduces snapshot recovery to a tractable linear algebra problem, solvable via Gaussian elimination.

Addressing the inherent linear correlation between gradients and snapshots in polynomial-DLA models, a dynamic mechanism called Dynamic Local Scrambling (DLS) is introduced to address the weak privacy breach. DLS not only prevents snapshot recovery attacks but also avoids the barren plateau problem associated with global random measurements. At the $t$-th training iteration, we apply a time-dependent unitary operator $W_t$ prior to measurement. Crucially, to preserve trainability, we restrict $W_t$ to be a tensor product of single-qubit 2-designs:
\begin{align}
    W_t = \bigotimes_{k=1}^n w_t^{(k)}, \quad w_t^{(k)} \sim \text{Haar}(SU(2))
\end{align}

This is physically equivalent to measuring a dynamic effective observable $O_{eff}^{(t)} = W_t^\dagger O W_t$.

The gradient $C_j^{(t)}$ observed by the adversary evolves into a stochastic linear combination of the snapshots:
\begin{align}
    C_j^{(t)} = \sum_{\alpha=1}^{\dim(\mathfrak{g})} \tilde{\chi}_{j\alpha}(W_t) [e_{snap}]_\alpha + \mathcal{R}_j(x, W_t) 
\end{align}

where the coefficients $\tilde{\chi}_{j\alpha}$ are random variables determined by the local unitaries $W_t$, and $\mathcal{R}_j$ represents the residual term arising from the operator escaping the DLA subspace. 

\textbf{Defense Analysis.} Weak Privacy Breach relies on the determinism of the gradient equation $C_j = \sum_\alpha A_{j\alpha} [e_{snap}]_\alpha$. The matrix $A$ is determined entirely by the ansatz parameters and the static measurement operator $O$. Instantaneous gradient generation follows DLA dynamics in our scheme. The adversary faces severe information scarcity. DLS introduces unknown time-varying unitary operators $W_t$. The effective measurement operator $O_{eff}^{(t)}$ becomes randomly unknown to the adversary. Construction of the correct coefficient matrix $A^{(t)}$ is impossible for the attacker. They are forced to construct an incorrect matrix based on static assumptions. This results in a mathematically inconsistent linear system. The adversary cannot utilize the DLA structure to recover snapshots even with a low DLA dimension.

\subsection{Truncated Chebyshev graph encoding (TCGE)}
The second layer of the defense architecture is presented in this section. Even if a weak privacy breach is assumed to have occurred, this layer ensures that recovering the original input $x$ from $e_{snap}$ remains computationally intractable. To address the strong privacy breach, the Truncated Chebyshev Graph Encoding (TCGE) is proposed. Unlike previous methods that compress high-order information into a single scalar rotation, TCGE employs a Chebyshev Tower strategy combined with graph state entanglement. The feature map $V(x)$ is defined as Equation \ref{eqvx}:
\begin{align}\label{eqvx}
    V(x) = U_{CZ}^\dagger \left( \bigotimes_{j=1}^n R_Y(\phi_j(x)) \right) U_{CZ} H^{\otimes n} 
\end{align}

TCGE incorporates two rigorous design constraints to balance privacy and trainability. The first constraint is the Chebyshev tower strategy. Inspired by the Fourier Tower architecture discussed in \cite{williams2023quantum}, we propose the Chebyshev Tower Strategy, which distributes hierarchical Chebyshev polynomials across the qubit register to maximize expressivity and nonlinearity. The rotation angle for the $j$-th qubit, $\phi_j(x)$, encodes a specific high-frequency component:
\begin{align}
    \phi_j(x) = 2 \cdot k_j \cdot \arccos(x_{mapped})
\end{align}

where $k_j$ represents the specific Chebyshev order assigned to qubit $j$ (e.g., $k=1$ for $q_0$, $k=2$ for $q_1$). This effectively injects the $k$-th order Chebyshev polynomial $T_k(x)$ directly into the state amplitude via the identity $R_Y(2k\theta)|0\rangle = \cos(k\theta)|0\rangle + \sin(k\theta)|1\rangle$.

The second constraint is the initialization of graph states. The operator $U_{CZ}$ is defined as a constant-depth sequence of Controlled-Phase (CZ) gates arranged in a nearest-neighbor linear topology. Unlike standard CNOT ladders, which act trivially on the uniform superposition basis created by the Hadamard layer ($H^{\otimes n}$), the CZ ladder effectively generates a 1D Linear Cluster State (Graph State). This shallow graph-state structure ensures that the initial state $\rho(x)$ maintains sufficient generalized purity with respect to the Poly-DLA basis, thus preventing state-induced trainability issues.

\textbf{Defense Analysis.} Strong privacy breach exploits the simplicity of snapshots as functions of input data. This simplicity often stems from the separability structure of encoded states. The graph connectivity in TCGE enforces global information diffusion. For any subsystem partition $J$, the reduced density matrix $\rho_J = \text{Tr}_{J^c}(\rho(x))$ becomes dependent on the full input vector $x = [x_1, \dots, x_d]$ due to the connectedness of the linear graph. Consequently, no independent subsystem $\rho_J(x_J)$ exists that depends only on a subset of variables $x_J$. This direct violation of the separability prerequisite renders the snapshot inversion algorithm inapplicable and effectively blocks the analytical inversion path.

For adversaries treating the VQC as a black-box function to be inverted via gradient descent, TCGE creates a rugged optimization landscape. The loss function $L(x')$ inherits frequency components up to order $K$ from the Chebyshev polynomials. According to Lipschitz continuity analysis, the query complexity to find an $\epsilon$-approximate solution scales with the Lipschitz constant $L \propto K$. By selecting $K = \mathcal{O}(\log n)$, we ensure that the density of local minima increases exponentially with the input dimension, forcing the adversary into a grid search regime with exponential complexity $\mathcal{O}((L/\epsilon)^d)$, which is computationally infeasible.

\subsection{Orthogonal decoupling of trainability and privacy}

Prior studies\cite{ragone2024lie,larocca2025barren} suggest that robust privacy necessitates highly expressive circuits. High expressivity corresponds to an exponentially large Dynamical Lie Algebra (DLA). Such immense dimensions mathematically result in barren plateaus\cite{cerezo2021cost}. Thus, the pursuit of privacy directly undermines trainability in traditional VQC frameworks.

The proposed architecture adopts the orthogonal decoupling strategy to overcome this barrier. The core concept involves separating the sources of privacy from the sources of trainability into distinct mathematical structures. The Orthogonal decoupling of trainability and privacy includes the following two aspects:
\begin{enumerate}
    \item \textbf{Source of Trainability}: The variance of the gradient is primarily determined by the dimension of the ansatz's DLA, $\dim(\mathfrak{g})$. By architecturally enforcing a Hamiltonian variational ansatz or a restricted Hardware Efficient Ansatz, we fix $\dim(\mathfrak{g}) = \text{poly}(n)$. Furthermore, the locality of DLS and the shallow graph-state structure of TCGE ensure that the signal overlap within the DLA subspace does not vanish exponentially.
    \item \textbf{Source of Privacy}: Privacy in our framework is not derived from the expressibility of the ansatz, but from the computational complexity of the input encoding and the stochastic obfuscation of gradients. Specifically, the Chebyshev Tower strategy in TCGE injects high-frequency nonlinearity to thwart inversion, while DLS provides dynamic randomness to block recovery.
\end{enumerate}

\subsubsection{Orthogonality of algebraic structures} 

Validating the decoupling strategy requires first proving that the externalized privacy mechanisms do not disrupt the polynomial DLA premise. It must be confirmed that the dimensionality of the parameter search space does not increase uncontrollably due to the introduced privacy components.

This is demonstrated by analyzing the algebraic composition of the loss function gradient. The partial derivative of the cost function with respect to a parameter $\theta_k$ is given by:
\begin{align}\label{eq_dialoss}
    \partial_k \mathcal{L} \propto \text{Tr}(O_{eff}^{(t)} \cdot \text{Ad}_{U}(\partial_k U \cdot U^\dagger) \cdot \rho_{TCGE}(x))
\end{align}

The central term in \ref{eq_dialoss}, the tangent vector field of the ansatz, strictly belongs to the Lie algebra $\mathfrak{g}$. Its dimension is determined entirely by the choice of ansatz generators. The use of the Hamiltonian Variational Ansatz in our scheme enforces a polynomial scaling law for $\dim(\mathfrak{g})$.

The privacy mechanisms modify the other two terms in the trace operation, namely the boundary conditions of evolution (effective measurement $O_{eff}^{(t)}$ and initial state $\rho_{TCGE}(x)$). Analysis shows these modifications solely alter the direction of projection onto the $\mathfrak{g}$ space. They absolutely do not alter the algebraic structure or dimension of the operator $\text{Ad}_{U}(\cdot)$ itself. The optimization path for parameters $\theta$ remains constrained within a low-dimensional submanifold. This proves that achieving privacy has not expanded the search space. The first necessary condition for trainability is satisfied.

\subsubsection{Preservation of signal variance} 
Guaranteeing a low-dimensional search space is insufficient to ensure trainability on its own. Prevention of "state-induced barren plateaus" is also necessary. This phenomenon occurs when the gradient signal slips into the null space of the DLA. It must be proven that the complex encoding does not cause signal loss.

The lower bound of the gradient variance is proportional to the squared magnitude of the projection of the initial state onto the algebra basis (generalized purity):
\begin{align}
    \text{Var}(\partial_k \mathcal{L}) \ge \frac{C_{\mathfrak{g}}}{\text{poly}(n)} \sum_{\alpha} (\text{Tr}(\rho_{TCGE}(x) B_\alpha))^2
\end{align}

Maintaining trainability requires that this generalized purity term in the numerator does not vanish exponentially.

The constant-depth constraint explicitly introduced in the TCGE encoding is the key physical mechanism guaranteeing this. Although the Chebyshev Tower strategy introduces high-degree polynomials into the state amplitudes, the underlying quantum circuit depth remains constant (specifically, depth-2 for the 1D linear graph state generation). Such constant-depth states do not reach a Haar-random distribution (which typically requires linear or polynomial depth). They retain significant signal overlap with the local Pauli operator basis. Consequently, the numerator in the variance formula remains of the order $\mathcal{O}(1)$. The gradient signal is physically preserved. This proves the second necessary condition for trainability.

\subsection{3-qubit VQC example}

We construct a  3-qubit  variational quantum circuit to demonstrate the implementation details of the DLPDA architecture. This comparison highlights the structural transition from a vulnerable baseline to the DyLoC model.

\subsubsection{Standard vulnerable VQC}
\begin{figure}
    \centering
    \includegraphics[width=1\linewidth]{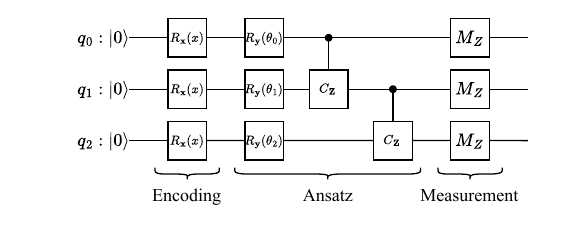}
    \caption{\textbf{Standard vulnerable VQC.} The baseline 3-qubit VQC architecture employs separable product encoding and static global measurements. This configuration establishes a deterministic linear mapping that exposes the model to algebraic snapshot recovery and inversion attacks.}
    \label{baseline_sample}
\end{figure}

The baseline model represents a typical configuration used in current (VQC) QML research. It is designed to satisfy the polynomial DLA constraint for trainability but lacks specific privacy defenses. The circuit topology consists of three sequential stages, as illustrated in Figure \ref{baseline_sample}. The encoding stage employs a product encoding strategy. The input scalar $x$ is mapped to local rotation angles via $R_X(x)$ gates on all qubits. The resulting state $|\psi_{in}\rangle$ is a product state. This structure satisfies the separability condition required by the inversion attack described in the snapshot inversion algorithm. The ansatz stage applies a single layer of a restricted Hardware Efficient Ansatz (HEA). It consists of parameterized $R_Y(\theta)$ rotations followed by a linear chain of $CZ$ gates for entanglement. The measurement stage performs a static global $Z$ projection where the target observable is $O = Z_0 Z_1 Z_2$.

The vulnerability of the baseline is twofold. The separability of the encoding allows the snapshot inversion problem to be decomposed into independent single-qubit subproblems. Furthermore, the static nature of the measurement operator $O$ creates a deterministic linear mapping between gradients and snapshots. This enables efficient snapshot recovery via the snapshot recovery algorithm.

\subsubsection{DyLoC-enhanced VQC}

\begin{figure}
    \centering
    \includegraphics[width=1\linewidth]{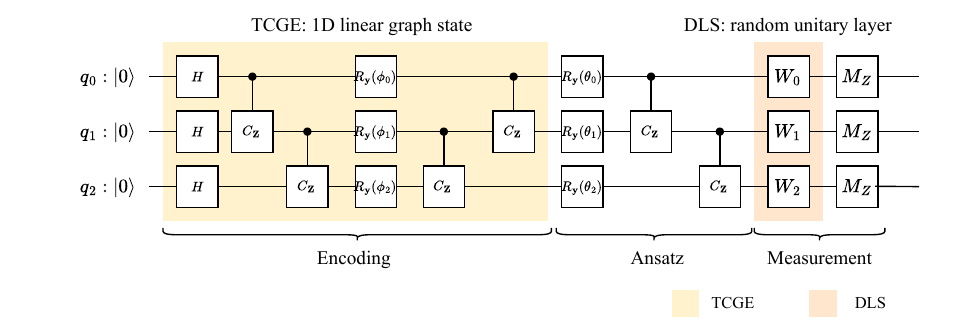}
    \caption{\textbf{DyLoC-enhanced VQC.} The DyLoC architecture integrates TCGE for enforced global entanglement and Dynamic Local Scrambling for gradient obfuscation. These components synergistically defend against algebraic attacks while preserving the trainability of the polynomial-DLA ansatz.}
    \label{DyLoC_sample}
\end{figure}

The secured model replaces the input and output interfaces with the proposed defense components, while the core ansatz remains unchanged to preserve the optimization landscape geometry. The circuit integrates TCGE and DLS, as illustrated in Figure \ref{DyLoC_sample}. 

TCGE executes the TCGE protocol. A layer of Hadamard gates ($H^{\otimes 3}$) is first applied to the vacuum state to create a uniform superposition $|+\rangle^{\otimes 3}$. A $CZ$ ladder ($0-1, 1-2$) is then applied. A $CZ$ gate is used instead of a CNOT gate because a CNOT gate acts trivially on the $|+\rangle$ basis. This step generates a multipartite 1D Linear graph State. Local rotations $R_Y(\phi_K(x))$ are subsequently applied, where the angle $\phi_K(x)$ is a $K$-th order Chebyshev polynomial. An inverse $CZ$ ladder is finally applied to propagate local phase injections into global multi-body correlations. The ansatz stage retains the same restricted HEA structure. 

DLS executes the Dynamic Local Scrambling. A random unitary layer $W_t$ is inserted prior to measurement. This layer consists of locally sampled single-qubit Haar random gates $w_t^{(0)} \otimes w_t^{(1)} \otimes w_t^{(2)}$. The physical measurement corresponds to the time-varying operator $O^{(t)}_{eff} = W_t^\dagger O W_t$.

\subsubsection{Security analysis}

The defense mechanism analysis confirms the security of DyLoC. The graph state initialization combined with the ladder structure ensures that the reduced density matrix of any subsystem depends on the full input vector $x$. This global dependency violates the separability assumption of the snapshot inversion algorithm. The high-frequency kernel forces inversion attacks into an exponential grid search regime, thereby achieving strong privacy. 

On the other hand, the introduction of the private random unitary $W_t$ randomizes the coefficients of the gradient equation. The adversary faces an inconsistent linear system. Snapshot recovery is mathematically blocked due to rank deficiency, thereby achieving weak privacy. The signal variance is preserved throughout the process. The shallow depth of TCGE maintains the overlap with the DLA basis. The locality of DLS prevents measurement-induced barren plateaus. The model remains trainable within the polynomial DLA subspace.

\section{Experiments}
\subsection{Experimental setup}
To comprehensively evaluate the performance of the proposed DyLoC architecture, numerical simulations were conducted focusing on three key dimensions: model trainability (utility), resilience against gradient-based recovery (weak privacy), and robustness against snapshot inversion (strong privacy). The experiments were implemented using the PennyLane quantum machine learning library.

The experiments utilized the Make-Moons dataset for non-linear binary classification, which consists of 150 samples with noise $\sigma=0.05$3. Features were mapped to the interval $[0, \pi]$ to match the optimal encoding range4. The following model architectures were evaluated:

\begin{enumerate}
    \item Standard VQC: Utilizes Pauli-X encoding and strongly entangling layers. This model represents the high-utility/low-privacy baseline.
    \item VQC + QDP: Adds Laplacian noise ($\lambda=0.15$) to gradients. This model represents the noise-based defense.
    \item DyLoC: Integrates Truncated Chebyshev Graph Encoding ($K=2$) and Dynamic Local Scrambling ($\delta\in[-0.3,0.3]$).
\end{enumerate}

\subsection{Trainability}

\begin{figure}
    \centering
    \includegraphics[width=0.8\linewidth]{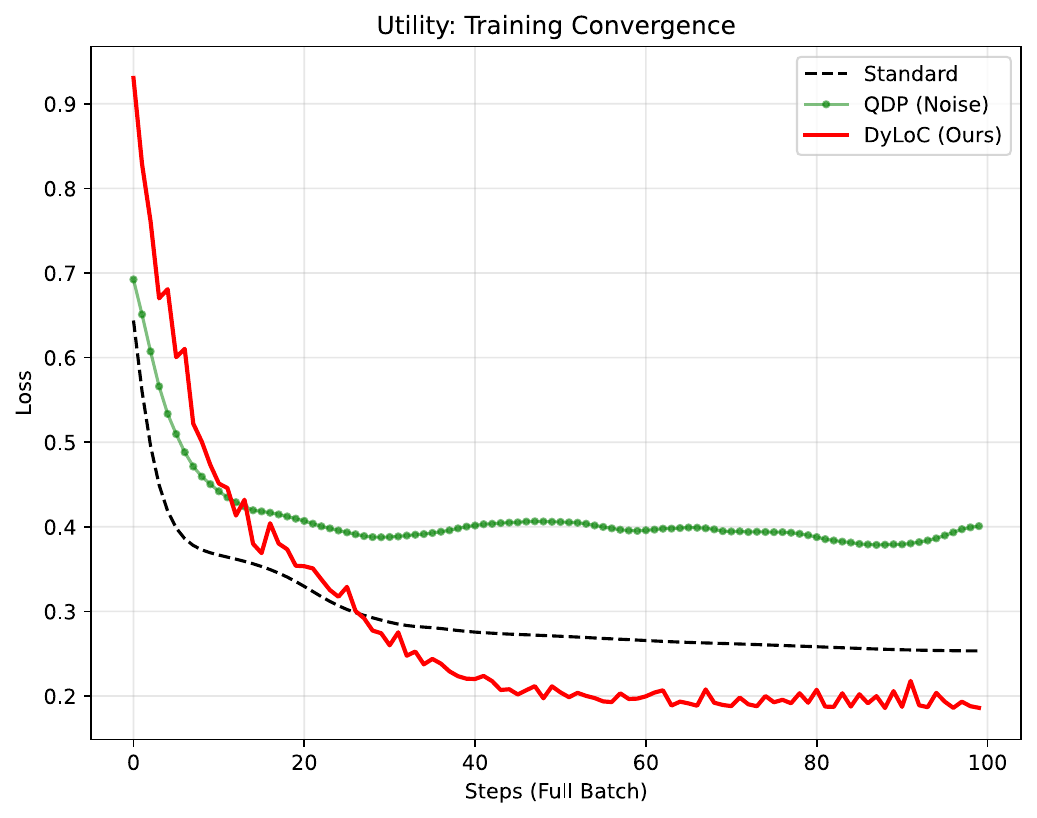}
    \caption{\textbf{Utility Comparison.} The training loss convergence demonstrates that DyLoC (red) maintains a stable descent trajectory comparable to the Standard baseline (black), whereas the QDP model (green) suffers from significant oscillation and fails to reach the optimal solution due to noise injection.}
    \label{training}
\end{figure}

The training convergence behavior is illustrated in Fig. \ref{training}. The Standard VQC (black dashed line) exhibits rapid convergence, descending to a loss below 0.3 within the first 40 steps and reaching a loss value of 0.25 at 100 steps. This trajectory demonstrates the ideal optimization path. Notably, the DyLoC model (red solid line) closely mirrors this trajectory. Despite the dynamic perturbations introduced by DLS, the loss curve maintains a consistent downward trend and converges to a final value ($\approx0.186$) lower than the standard baseline. These results confirm that the orthogonal decoupling strategy effectively isolates the gradient signal and prevents optimization collapse11. In contrast, the VQC + QDP model (green line) suffers from observable instability. The noise injection causes the loss to oscillate around a higher plateau ($\approx0.4$) and fails to reach the optimal solution found by the other two models.

\subsection{Evaluation of weak privacy breach}
To quantitatively evaluate the protection effect against weak privacy leakage, the Weak Privacy MSE is defined as Definition \ref{weakMSE}.

\begin{definition}[Weak Privacy MSE]\label{weakMSE}
 Let $g_{real}^{(t)} = \nabla_\theta \mathcal{L}(W_t)$ be the gradient generated by DyLoC at step $t$, and $g_{static} = \nabla_\theta \mathcal{L}(I)$ be the gradient estimated by the adversary using the static ansatz. The Weak Privacy MSE is defined as:
\begin{align}
    \text{MSE}_{\text{weak}} = \frac{1}{D} \| g_{real}^{(t)} - g_{static} \|^2
\end{align}
 
 where $D$ is the number of parameters. 
\end{definition}

A higher $\text{MSE}_{\text{weak}}$ indicates stronger gradient obfuscation, implying that the adversary's linear system for snapshot recovery is mathematically inconsistent and rank-deficient. This metric quantifies the discrepancy between the true gradient observed from the dynamic system and the theoretical gradient derived by an adversary assuming a static model.

\begin{figure}
    \centering
    \includegraphics[width=0.9\linewidth]{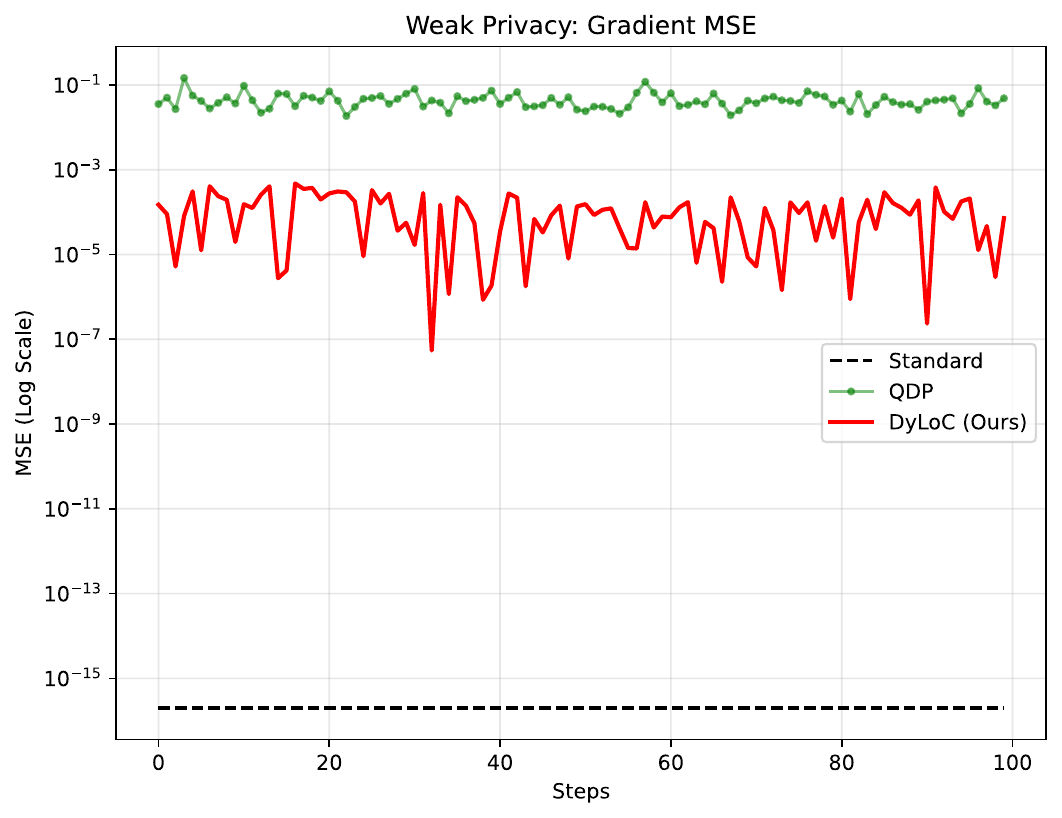}
    \caption{\textbf{Weak Privacy Evaluation.} The gradient reconstruction MSE indicates that DyLoC imposes a persistent structural mismatch for the adversary, maintaining a high error magnitude ($10^{-2}\sim 10^{-3}$) compared to the negligible error ($10^{-16}$) of the Standard baseline.}
    \label{weak_privacy}
\end{figure}

Fig. \ref{weak_privacy} presents the Mean Squared Error (MSE) of the reconstructed gradients. The Standard VQC curve flatlines at the numerical precision floor ($10^{-16}$), indicating total exposure to gradient inversion attacks.

The DyLoC curve (red) maintains a high MSE, fluctuating between $10^{-3}$ and $10^{-2}$. This signifies a persistent structural mismatch between the adversary's static assumption and the true dynamic process. While QDP (green) achieves a slightly higher MSE due to additive noise, this advantage is marginal compared to the severe utility cost observed in Fig. \ref{training}. DyLoC achieves robust obfuscation—raising the reconstruction error by 13 orders of magnitude compared to the baseline—without sacrificing convergence.

\subsection{Evaluation of strong privacy breach}
To quantitatively evaluate the robustness against snapshot inversion, the Strong Privacy MSE is defined as Definition \ref{weakMSE}.

\begin{definition}[Strong privacy MSE ]\label{strongMSE}
This metric measures the success of an inversion attack by calculating the Euclidean distance between the ground-truth input data $x_{true}$ and the adversary's reconstructed input $x_{adv}$ after optimization.
\begin{align}
  \text{MSE}_{\text{strong}} = \frac{1}{N} \| x_{true} - x_{adv} \|^2  
\end{align}

where $N$ is the input dimension.
\end{definition}

 A higher $\text{MSE}_{\text{strong}}$ indicates a successful defense, showing that the adversary failed to converge to the correct input data due to the rugged loss landscape created by TCGE.

\begin{figure}
    \centering
    \includegraphics[width=0.9\linewidth]{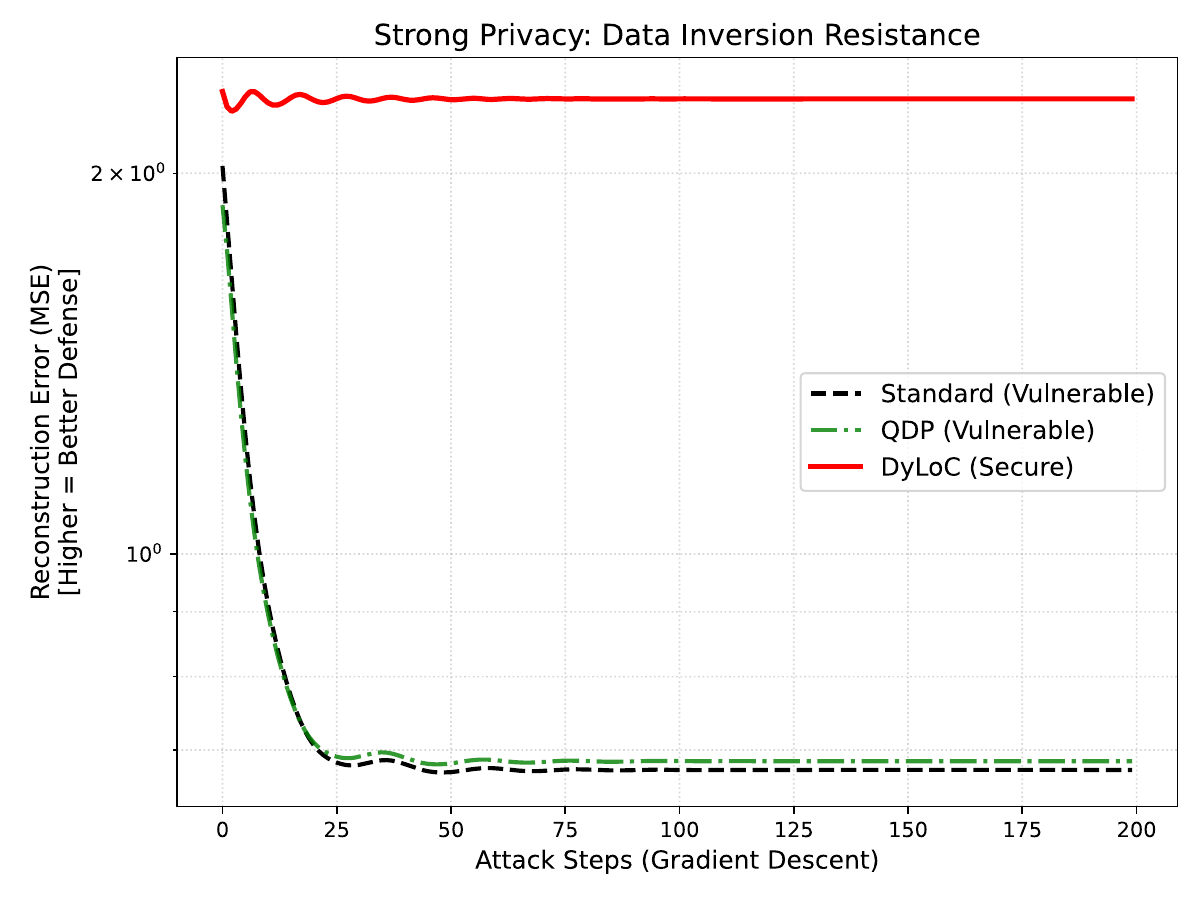}
    \caption{\textbf{Strong Privacy Evaluation.} DyLoC architecture effectively traps the adversary at the initialization point with a high reconstruction error ($\text{MSE} > 2.0$), while Standard and QDP models are rapidly inverted to almost zero error within 50 iterations.}
    \label{strong_privacy}
\end{figure}

The robustness against snapshot inversion is quantified in Fig. \ref{strong_privacy}. This experiment simulates an adversary utilizing gradient-based optimization (Adam) to reconstruct the input data $x$ from a leaked snapshot. The attack is initialized with a random guess far from the target to test global convergence capabilities.

In Fig. \ref{strong_privacy}, the Standard VQC (black dashed line) and VQC + QDP (green dot-dash line) exhibit nearly identical convergence behaviors. The reconstruction error for both models decays monotonically and exponentially. It reaches a negligible level ($\text{MSE} < 0.2$) within 50 iterations. This result confirms that noise injection during training (QDP) does not alter the geometric smoothness of the model's landscape during the inference phase. Standard architectures are thus proven to be highly vulnerable to inversion attacks.

In sharp contrast, the DyLoC model (red solid line) demonstrates complete resistance, as shown in Fig. \ref{strong_privacy}. The reconstruction error fails to descend. It remains stagnant at the magnitude of the initial guess error ($\text{MSE} > 2.0$). No effective gradient descent trajectory is established throughout the attack window.

The phenomenon represents a state of Gradient Masking. TCGE creates a loss landscape characterized by extreme ruggedness and dense local minima. At the adversary's initialization point, the gradients derived from the DyLoC circuit are either vanishing or point in non-descent directions due to the high-frequency oscillations of the Chebyshev tower. The adversary is effectively trapped at the initialization point, unable to extract any meaningful information to navigate towards the true input. Consequently, strong privacy is preserved with near-ideal efficacy.

\begin{figure}
    \centering
    \includegraphics[width=1\linewidth]{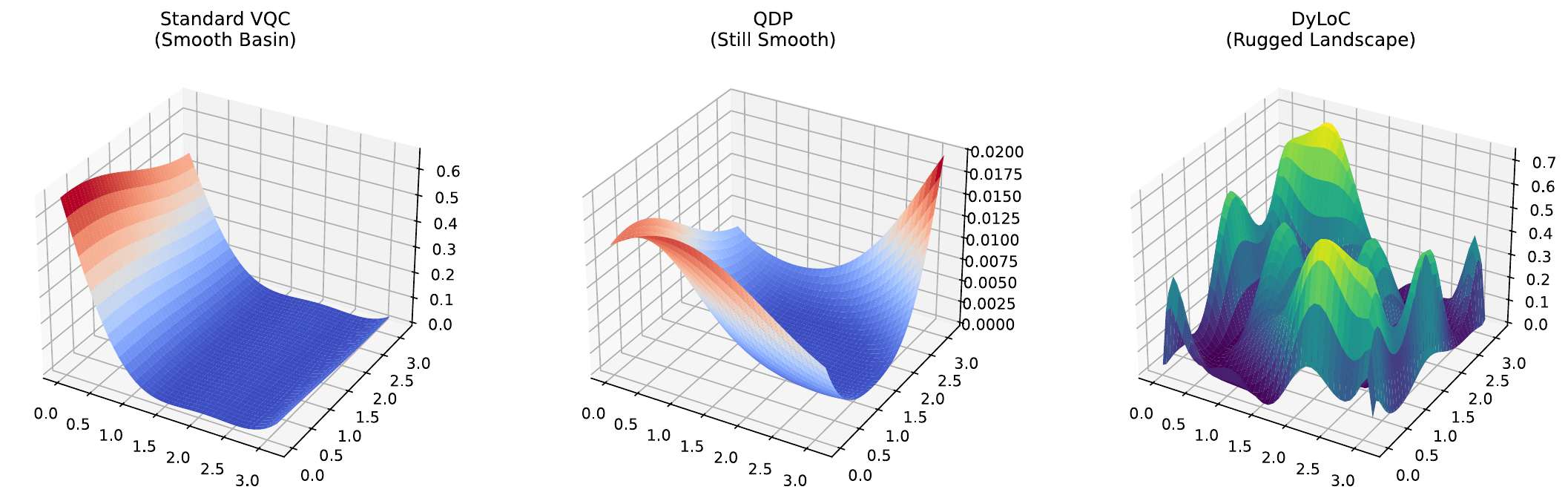}
    \caption{\textbf{Inversion Landscape Comparison.} The Standard VQC (left)and QDP (middle) display a smooth convex basin facilitating easy attack convergence, whereas the DyLoC landscape (right) exhibits extreme ruggedness induced by TCGE that effectively traps gradient-based attackers in local minima.}
    \label{landscape}
\end{figure}

The effectiveness of DyLoC in protecting against strong privacy breach is further explained by the loss landscapes in Fig. \ref{landscape}. The Standard VQC (left) and QDP (middle) present a smooth, convex basin that guides optimization directly to the global minimum. In stark contrast, the DyLoC landscape (right) is characterized by a rugged, multi-modal topology filled with local minima. The complex geometry creates a prohibitive barrier for inversion attacks, which is induced by the graph-state entanglement and Chebyshev tower structure. 

\section{Conclusion}
This paper addressed the key challenge of securing Variational Quantum Circuits against algebraic privacy attacks without compromising trainability. The DyLoC architecture was proposed to secure Variational Quantum Circuits against algebraic privacy attacks while preserving trainability. By implementing an orthogonal decoupling strategy, the scheme separated privacy sources from the algebraic structure of the ansatz. The Truncated Chebyshev Graph Encoding defeated snapshot inversion through high-frequency nonlinearity and graph-state entanglement. Concurrently, the Dynamic Local Scrambling mitigated snapshot recovery by dynamically obfuscating gradient linearity. Experimental validation confirmed that the framework blocked algebraic attacks with high reconstruction error and maintained convergence comparable to unprotected baselines. Future research investigates hardware-efficient implementations on specific topological constraints and extension to other quantum neural network architectures.

\bibliographystyle{IEEEtran}
\bibliography{ref}

@article{heredge2025characterizing,
  title={Characterizing privacy in quantum machine learning},
  author={Heredge, Jamie and Kumar, Niraj and Herman, Dylan and Chakrabarti, Shouvanik and Yalovetzky, Romina and Sureshbabu, Shree Hari and Li, Changhao and Pistoia, Marco},
  journal={npj Quantum Information},
  volume={11},
  number={1},
  pages={80},
  year={2025},
  publisher={Nature Publishing Group UK London}
}

@inproceedings{ju2024harnessing,
  title={Harnessing inherent noises for privacy preservation in quantum machine learning},
  author={Ju, Keyi and Qin, Xiaoqi and Zhong, Hui and Zhang, Xinyue and Pan, Miao and Liu, Baoling},
  booktitle={ICC 2024-IEEE International Conference on Communications},
  pages={1121--1126},
  year={2024},
  organization={IEEE}
}

@article{cerezo2021cost,
  title={Cost function dependent barren plateaus in shallow parametrized quantum circuits},
  author={Cerezo, Marco and Sone, Akira and Volkoff, Tyler and Cincio, Lukasz and Coles, Patrick J},
  journal={Nature communications},
  volume={12},
  number={1},
  pages={1791},
  year={2021},
  publisher={Nature Publishing Group UK London}
}

@article{ragone2024lie,
  title={A Lie algebraic theory of barren plateaus for deep parameterized quantum circuits},
  author={Ragone, Michael and Bakalov, Bojko N and Sauvage, Fr{\'e}d{\'e}ric and Kemper, Alexander F and Ortiz Marrero, Carlos and Larocca, Mart{\'\i}n and Cerezo, Marco},
  journal={Nature Communications},
  volume={15},
  number={1},
  pages={7172},
  year={2024},
  publisher={Nature Publishing Group UK London}
}

@article{larocca2025barren,
  title={Barren plateaus in variational quantum computing},
  author={Larocca, Martin and Thanasilp, Supanut and Wang, Samson and Sharma, Kunal and Biamonte, Jacob and Coles, Patrick J and Cincio, Lukasz and McClean, Jarrod R and Holmes, Zo{\"e} and Cerezo, Marco},
  journal={Nature Reviews Physics},
  pages={1--16},
  year={2025},
  publisher={Nature Publishing Group UK London}
}

@article{williams2023quantum,
  title={Quantum chebyshev transform: Mapping, embedding, learning and sampling distributions},
  author={Williams, Chelsea A and Paine, Annie E and Wu, Hsin-Yu and Elfving, Vincent E and Kyriienko, Oleksandr},
  journal={arXiv preprint arXiv:2306.17026},
  year={2023}
}

@article{mcclean2018barren,
  title={Barren plateaus in quantum neural network training landscapes},
  author={McClean, Jarrod R and Boixo, Sergio and Smelyanskiy, Vadim N and Babbush, Ryan and Neven, Hartmut},
  journal={Nature communications},
  volume={9},
  number={1},
  pages={4812},
  year={2018},
  publisher={Nature Publishing Group UK London}
}

@article{holmes2022connecting,
  title={Connecting ansatz expressibility to gradient magnitudes and barren plateaus},
  author={Holmes, Zo{\"e} and Sharma, Kunal and Cerezo, Marco and Coles, Patrick J},
  journal={PRX quantum},
  volume={3},
  number={1},
  pages={010313},
  year={2022},
  publisher={APS}
}

@article{du2021quantum,
  title={Quantum noise protects quantum classifiers against adversaries},
  author={Du, Yuxuan and Hsieh, Min-Hsiu and Liu, Tongliang and Tao, Dacheng and Liu, Nana},
  journal={Physical Review Research},
  volume={3},
  number={2},
  pages={023153},
  year={2021},
  publisher={APS}
}

@article{watkins2023quantum,
  title={Quantum machine learning with differential privacy},
  author={Watkins, William M and Chen, Samuel Yen-Chi and Yoo, Shinjae},
  journal={Scientific Reports},
  volume={13},
  number={1},
  pages={2453},
  year={2023},
  publisher={Nature Publishing Group UK London}
}

@article{qvarfort2025solving,
  title={Solving Quantum Dynamics with a Lie-Algebra Decoupling Method},
  author={Qvarfort, Sofia and Pikovski, Igor},
  journal={PRX Quantum},
  volume={6},
  number={1},
  pages={010201},
  year={2025},
  publisher={APS}
}

@article{wiersema2020exploring,
  title={Exploring entanglement and optimization within the Hamiltonian variational ansatz},
  author={Wiersema, Roeland and Zhou, Cunlu and de Sereville, Yvette and Carrasquilla, Juan Felipe and Kim, Yong Baek and Yuen, Henry},
  journal={PRX quantum},
  volume={1},
  number={2},
  pages={020319},
  year={2020},
  publisher={APS}
}

@inproceedings{broadbent2009universal,
  title={Universal blind quantum computation},
  author={Broadbent, Anne and Fitzsimons, Joseph and Kashefi, Elham},
  booktitle={2009 50th annual IEEE symposium on foundations of computer science},
  pages={517--526},
  year={2009},
  organization={IEEE}
}

@article{li2021quantum,
  title={Quantum federated learning through blind quantum computing},
  author={Li, Weikang and Lu, Sirui and Deng, Dong-Ling},
  journal={Science China Physics, Mechanics \& Astronomy},
  volume={64},
  number={10},
  pages={100312},
  year={2021},
  publisher={Springer}
}

@article{kumar2023expressive,
  title={Expressive variational quantum circuits provide inherent privacy in federated learning},
  author={Kumar, Niraj and Heredge, Jamie and Li, Changhao and Eloul, Shaltiel and Sureshbabu, Shree Hari and Pistoia, Marco},
  journal={arXiv preprint arXiv:2309.13002},
  year={2023}
}

@article{biamonte2017quantum,
  title={Quantum machine learning},
  author={Biamonte, Jacob and Wittek, Peter and Pancotti, Nicola and Rebentrost, Patrick and Wiebe, Nathan and Lloyd, Seth},
  journal={Nature},
  volume={549},
  number={7671},
  pages={195--202},
  year={2017},
  publisher={Nature Publishing Group UK London}
}

@article{preskill2018quantum,
  title={Quantum computing in the NISQ era and beyond},
  author={Preskill, John},
  journal={Quantum},
  volume={2},
  pages={79},
  year={2018},
  publisher={Verein zur F{\"o}rderung des Open Access Publizierens in den Quantenwissenschaften}
}

@article{cerezo2021variational,
  title={Variational quantum algorithms},
  author={Cerezo, Marco and Arrasmith, Andrew and Babbush, Ryan and Benjamin, Simon C and Endo, Suguru and Fujii, Keisuke and McClean, Jarrod R and Mitarai, Kosuke and Yuan, Xiao and Cincio, Lukasz and others},
  journal={Nature Reviews Physics},
  volume={3},
  number={9},
  pages={625--644},
  year={2021},
  publisher={Nature Publishing Group UK London}
}

\end{document}